# Generative Adversarial Network-Based Sinogram Super-Resolution for Computed Tomography Imaging

Chao Tang, Wenkun Zhang, Linyuan Wang, Ailong Cai, Ningning Liang, Lei Li, and Bin Yan

Henan Key Laboratory of Imaging and Intelligent Processing, PLA Strategic Support Force Information Engineering University, Zhengzhou, China

E-mail: ybspace@hotmail.com

**Abstract**

Compared with the conventional 1×1 acquisition mode of projection in computed tomography (CT) image reconstruction, the 2×2 acquisition mode improves the collection efficiency of the projection and reduces the X-ray exposure time. However, the collected projection based on the 2×2 acquisition mode has low resolution (LR) and the reconstructed image quality is poor, thus limiting the use of this mode in CT imaging systems. In this study, a novel sinogram-super-resolution generative adversarial network (SSR-GAN) model is proposed to obtain high-resolution (HR) sinograms from LR sinograms, thereby improving the reconstruction image quality under the 2×2 acquisition mode. The proposed generator is based on the residual network for LR sinogram feature extraction and super-resolution (SR) sinogram generation. A relativistic discriminator is designed to render the network capable of obtaining more realistic SR sinograms. Moreover, we combine the cycle consistency loss, sinogram domain loss, and reconstruction image domain loss in the total loss function to supervise SR sinogram generation. Then, a trained model can be obtained by inputting the paired LR/HR sinograms into the network. Finally, the classic FBP reconstruction algorithm is used for CT image reconstruction based on the generated SR sinogram. The qualitative and quantitative results of evaluations on digital and real data illustrate that the proposed model not only obtains clean SR sinograms from noisy LR sinograms but also outperforms its counterparts.

## 1. Introduction

Computed tomography (CT) technology has been widely applied to medical imaging [1], industrial non-destructive testing [2], and safety inspection [3]. In practice, the fidelity of CT image features may be enhanced by high-resolution (HR) imaging, which can better guide high-accuracy radiotherapy [4]. However, the resolution of reconstruction images in a CT imaging system is constrained by factors such as the detector element pitch, the focal spot size of the X-ray, and image reconstruction algorithms. In some important applications, such as early tumor characterization and coronary artery analysis, the inherent resolution of modern CT imaging systems is still considerably lower than the ideal resolution [5]. More sophisticated hardware components can be used to improve the resolution, e.g., a detector element with a small pitch, an X-ray tube with a fine focal spot size, or better mechanical precision for CT scanning. However, these hardware-oriented methods will increase the CT system costs and radiation dose and compromise imaging speeds. Thus, developing methods to obtain higher resolution images without changing the original hardware system has gained significant research attention [6].

In particular, some flat panel detectors such as Varian 2520D and 4030CB are equipped with two acquisition modes, namely 1×1 and 2×2 acquisition modes (www.varian.com), in most cone-beam CT imaging systems. The 2×2 acquisition mode enables fast scanning and improves projection collection efficiency because this mode has a higher acquisition frame rate. It can also significantly reduce the X-ray exposure time and the radiation dose under the condition of the same signal-to-noise ratio as that in the 1×1 acquisition mode. However, the 2×2 acquisition mode is rarely adopted by researchers because it obtains a low-resolution (LR) projection, resulting in the low quality of the reconstructed image. One possible strategy to improve the quality of the CT image under the 2×2 acquisition mode is to develop a method of super-resolution (SR) of the projection in order to obtain HR projection from LR projection.

In recent years, various SR methods have been proposed for improving the image resolution. These methods can be broadly divided into model- and learning-based approaches. The model-based approaches explicitly simulate the process of actually obtaining LR images and regularize the reconstruction on the basis of the prior information of image data [7-12]. These models can be broadly described as follows:

$$y^* = \arg\min_y \frac{1}{2} \| D_\downarrow B * y - x \|_2^2 + \lambda R(y),  \quad (1)$$

where $y$ denotes the HR image, $x$ denotes the LR image, $*$ denotes the convolution operation, and $D_\downarrow B$ denotes the down-sampling and blurring operations. The functional $R(\cdot)$ denotes the prior information, and multiplying it with parameter $\lambda$ sets the relative weights between two penalties. These methods promise excellent reconstruction image quality under the condition that the image priors are effectively mined. Learning-based approaches can recover the missing high-frequency image details by learning a nonlinear LR–HR mapping based on the training datasets. The approaches based on sparse representation also attract an increasing interest because it shows good performance in suppressing noise and artifacts [13-16]. Yang et al. [13] proposed a sparse representation-based SR reconstruction method, which needs to learn only two concise dictionaries and can obtain excellent results in terms of the peak signal-to-noise ratio (PSNR). Moreover, Zhang et al. [14] developed a patch-based sparse representation SR method for lung CT images and obtained good performance in artifact suppression. Although these methods can enhance the quality of images, the results may have blocky artifacts and lose some subtleties.

Recently, deep learning (DL) techniques have been developed for computer vision tasks [17]. In natural image SR tasks, convolutional neural networks (CNNs) have shown excellent performance [18-21]. In 2014, a three-layer CNN, named SRCNN, was first proposed by Dong et al. [18]. It learns a nonlinear LR–HR mapping in an end-to-end manner, exhibiting superior performance compared to traditional methods. In 2016, Wang [22] reported that DL technology is expected to promote further development of CT imaging technology. Then, DL-based SR methods were explored by researchers for CT imaging [23-27]. Yu et al. [23] proposed two advanced CNN-based models to preserve high-frequency textures. Park et al. [24] proposed a CNN-based approach that consists of a contracting path and a symmetric expanding path to produce SR CT images. More recently, You et al. [27] proposed a semi-supervised network model based on generative adversarial networks (GANs) to generate HR CT images. These methods mainly address the super-resolution problem of reconstructed image; however, the approaches of super-resolution in the projection domain are not fully explored. In the 2×2 acquisition mode, the LR property of raw projection data is extended to the image domain in the reconstruction process, which intrinsically restricts the above algorithms to obtain high-quality results. To improve the quality of CT reconstruction images under the 2×2 acquisition mode, adding the effective information of the projection domain by sinogram SR is a useful approach.

The main challenges in the sinogram SR problem are as follows. First, the spatial variations, correlations, and statistical properties of the sinograms and CT images differ, limiting the existing SR methods to some extent. Second, since the blurring operations and down-sampling are ill-posed and coupled in the process of projection acquisition, it is difficult to build a numerical model to accurately depict the projection degradation process. Finally, raw projection data are sensitive, with a small change in them having a significant effect on the reconstruction image quality; therefore, there is a requirement for higher accuracy of the SR algorithm. To address these limitations, we propose a GAN-based sinogram SR method in this study. Inspired by the cycle-GAN [28, 29], we propose a novel sinogram-super-resolution GAN model (SSR-GAN). The cycle-consistent structure design of the SSR-GAN impels the network to learn the mapping relationship between the LR and HR sinograms. In addition, we design a generator based on the residual CNN, which can weaken the gradient vanishing problem. Further, we utilize the relativistic discriminator, which facilitates the generator to produce HR sinogram more efficiently. In particular, we design a filtered-back-projection (FBP) module to achieve cross-domain error backpropagation. The generated SR sinogram is more accurate under the constraint of the weighted image domain loss. The LR–HR mapping is learned by training the SSR-GAN with paired LR/HR sinograms. Then, the classic FBP algorithm [30] is used to obtain a high-quality CT image from the SR sinogram. The evaluation results demonstrate that our proposed model exhibits excellent sinogram SR performance and effectively reduces artifacts in CT images.

The rest of this paper is organized as follows. In section 2, we review the SR problems in the imaging field and introduce the network architecture of the proposed SSR-GAN model. The experimental designs and quantitative evaluations on simulated and real data are reported in sections 3 and 4, respectively. Finally, the results and related issues as well as future research direction are discussed in section 5.

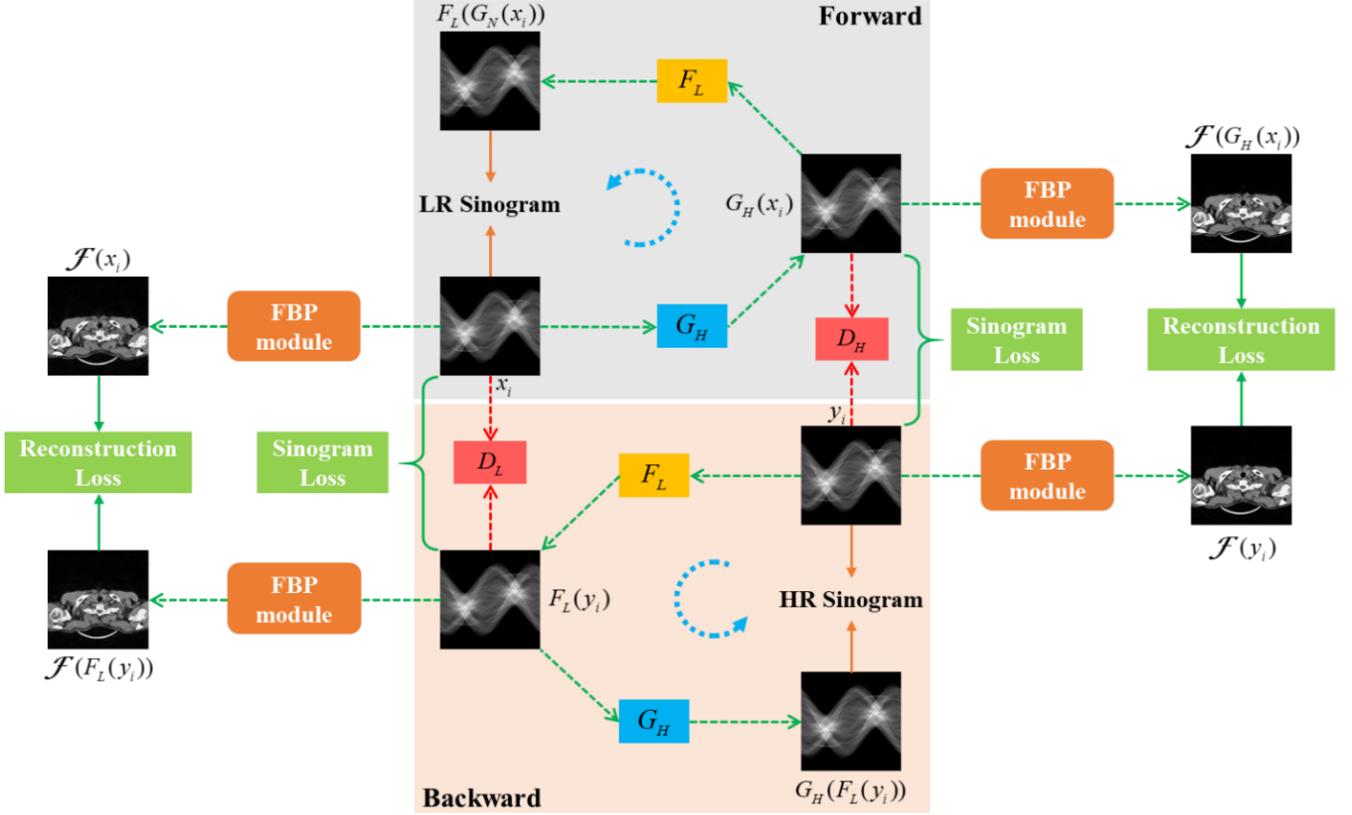

**Figure 1.** Schematic of the SSR-GAN framework. Our model contains two generators $G_H$ and $F_L$, and the corresponding discriminators $D_H$ and $D_L$ respectively, where $x_i$ denotes the LR sinogram and $y_i$ is the HR sinogram. The yellow rounded rectangles are the FBP modules used to calculate the reconstruction image loss.

## 2. Method

### 2.1 SR Problem Statement

In general, the conventional formulation of the ill-posed linear image SR problem [16] can be formulated as

$$x = D_\downarrow B * y + n, \qquad (2)$$

where $n$ denotes the noise and other factors. It is worth noting that both the down-sampling and blurring operations are nonlinear in practice. In addition, the nonlinearity exists in nonmodeled factors.

Our purpose is to convert LR sinograms obtained by the 2×2 acquisition mode to HR sinograms. However, the sinogram SR task is a nonlinear and ill-posed problem, and there are many limitations in the sinogram domain. In recent years, DL-based methods have provided new ideas for the nonlinear and ill-posed problem. These methods use feed-forward CNNs to learn the mapping from input data to label data by optimizing parameters $\theta$. The process can be written as:

$$\hat{y} = \Phi_\theta(x). \qquad (3)$$

In the sinogram SR problem, to generate an excellent SR sinogram $\hat{y}$, an appropriate network framework $\Phi_\theta$ must be designed and the loss function should be specified to facilitate $\Phi_\theta$ to obtain an SR sinogram based on the training datasets, so that

$$\hat{\theta} = \arg\min_\theta \sum_i \mathcal{L}(\hat{y}_i(\theta), y_i), \qquad (4)$$

where $(x_i, y_i)$ are the paired LR/HR sinograms for training.

*2.2 Sinogram SR Model*

An overview of the proposed network model is shown in Figure 1. The network model contains forward and backward cycles. In the forward cycle, the LR sinogram $x_i$ is input to the trained generator $G_H$ to obtain an SR sinogram $G_H(x_i)$. The trained generator $F_L$ translates the SR sinogram $G_H(x_i)$ back to the original LR sinogram. Similarly, in the backward cycle, the HR sinogram $y_i$ is input to the trained generator $F_L$ to obtain an LR sinogram $F_L(y_i)$. The trained generator $G_H$ translates the LR sinogram $F_L(y_i)$ back to the original HR sinogram. In addition, the discriminators $D_H$ and $D_L$ intend to identify that the sample is from real sinogram rather than generating sinogram. In the training process of the network, the generators $G_H$ and $F_L$ attempt to generate sinograms that are not easily distinguishable by the discriminators. Given an LR sinogram $x$, $G_H$ attempts to generate a new sinogram $G_H(x)$ highly similar to the corresponding HR sinogram $y$ so as to fool $D_H$. The key idea of the "game" process is to perform joint/alternative training on the generator and discriminator to collaboratively improve the performance of the network. Therefore, we can describe the optimization problem as

$$\min_{G_H, F_L} \max_{D_H, D_L} \mathcal{L}_{GAN}(G_H, D_H) + \mathcal{L}_{GAN}(F_L, D_L). \quad (5)$$

To promote high-precision SR sinogram generation, our proposed network model combines the following loss functions.

1) **Relativistic Adversarial GAN Loss**: Adversarial losses [31] are employed to prompt the generated sinograms to obey the empirical distributions in the source and target domains. Inspired by the relativistic GAN [32], we replace the standard discriminator with a relativistic discriminator, which predicts the probability that a real HR sinogram $y$ is relatively more realistic than a fake sinogram. As shown in Figure 2, the standard discriminator can be expressed as $D_H(y) = \sigma(C(y))$, where $\sigma$ denotes the sigmoid function and $C(y)$ denotes the non-transformed output of discriminator. The formulation of the relativistic discriminator is $D_H(y, G_H(x)) = \sigma(C(y) - \mathbb{E}\, C(G_H(x)))$, where $\mathbb{E}\,\cdot$ denotes the expectation operator and $x$ is the LR sinogram.

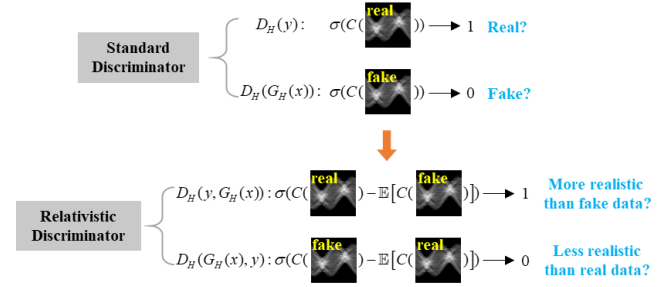

**Figure 2.** Difference between the standard discriminator and the relativistic discriminator.

Thus, the adversarial loss can be defined as

$$\mathcal{L}_{GAN}^{R}(G_H, D_H) = \mathbb{E}_{x,y \sim P_{data}(x,y)}\left[\log D_H(y, G_H(x))\right] + \mathbb{E}_{x,y \sim P_{data}(x,y)}\left[\log(1 - D_H(G_H(x), y))\right], \quad (6)$$

$$\mathcal{L}_{GAN}^{R}(G_L, D_L) = \mathbb{E}_{x,y \sim P_{data}(x,y)}\left[\log D_L(x, G_L(y))\right] + \mathbb{E}_{x,y \sim P_{data}(x,y)}\left[\log(1 - D_L(G_L(y), x))\right], \quad (7)$$

where $P_{data}$ denotes the sinogram data distribution.

2) **Cycle Consistency Loss**: To ensure that the learned mapping transforms the source input successfully to the target output, we utilize the cycle consistency loss to constrain $x$ and $F_L(G_H(x))$ in the forward cycle as well as $y$ and $G_H(F_L(y))$ in the backward cycle, as expressed below:

$$\mathcal{L}_{cyc}(G_H, F_L) = \mathbb{E}_{x \sim P_{data}(x)}\left[\|F_L(G_H(x)) - x\|_1\right] + \mathbb{E}_{y \sim P_{data}(y)}\left[\|G_H(F_L(y)) - y\|_1\right], \quad (8)$$

where $\|\cdot\|_1$ is the $L_1$ norm. The cycle consistency loss urges $F_L(G_H(x)) \approx x$ and $G_H(F_L(y)) \approx y$ to prevent the degeneracy in adversarial learning [33].

3) **Sinogram Domain Loss**: For the processing of sinograms, the accuracy of the generated sinogram is critical. The cycle consistency loss can only provide an indirect supervision, which is not sufficient to guarantee the accuracy of the final result. Therefore, the direct constraints to the generators are necessary, especially to $G_H$, which directly produces the desired SR sinograms. In our sinogram SR task, the outcomes and labels should be matched in each pixel. Previous methods have illustrated that the MAE loss is beneficial for image restoration in the pixel level [34, 35]. Therefore, we constructed the sinogram domain loss function based on MAE loss, as follows:

$$\mathcal{L}_{sino}(G_H, F_L) = \mathbb{E}_{x,y \sim P_{data}(x,y)}\left[\|F_L(y) - x\|_1\right] + \mathbb{E}_{x,y \sim P_{data}(x,y)}\left[\|G_H(x) - y\|_1\right], \quad (9)$$

4) **Reconstruction Image Domain Loss**: For the sinogram SR task, the sinogram data are sensitive. Small errors in the generated sinogram may cause critical artifacts in the reconstructed image. In our study, to address this problem, the image domain loss is added in the proposed network model. The additional image domain loss can result in more realistic generated sinograms, achieved by the ASTRA-Toolbox [36]. In addition, this study uses the MAE loss to compare the reconstructed images of the generated and real sinograms, as described below:

$$\mathcal{L}_{image}(G_H, F_L) = \mathbb{E}_{x,y \sim P_{data}(x,y)} \left[ \|\mathcal{A}(F_L(y)) - \mathcal{A}(x)\|_1 \right] + \mathbb{E}_{x,y \sim P_{data}(x,y)} \left[ \|\mathcal{A}(G_H(x)) - \mathcal{A}(y)\|_1 \right], \quad (10)$$

where $\mathcal{A}$ denotes the filtered-back-projection operation.

Therefore, the proposed model is trained to minimize the following total loss function:

$$\begin{aligned}\mathcal{L}_{SSRGAN} &= \mathcal{L}^R_{GAN}(G_H, D_H) + \mathcal{L}^R_{GAN}(G_L, D_L) \\ &+ \lambda_1 \mathcal{L}_{cyc}(G_H, F_L) + \lambda_2 \mathcal{L}_{sino}(G_H, F_L) \\ &+ \lambda_3 \mathcal{L}_{image}(G_H, F_L),\end{aligned} \quad (11)$$

where $\lambda_1$, $\lambda_2$, and $\lambda_3$ are the weight parameters for balancing different penalties.

*2.3 Network Architecture*

In the proposed method, the generator and discriminator of our network model play crucial roles. The specific architecture of the model is described as follows.

1) **Generator**: As shown in Figure 3(a), the overall architecture of the generator is similar to the traditional U-Net [37], except that, instead of the encoding–decoding structure, our generator is composed of three parts, namely the encoder, decoder, and transformation modules. In addition, the architecture of the encoder and decoder is more lightweight. As shown in Figure 3(b), the transformation module is a residual network (ResNet) [38], containing six ResNet blocks. The size of input sinograms in our generator is $i \times 512 \times 512 \times 1$, where $i$ is the batch size. In our generator, the encoder extracts features from the input sinogram to obtain the feature vector, with size $128 \times 128 \times 256$. Then, the transformation module converts the feature vector extracted from source domain X to target domain Y. The decoder restores the feature vector to low-level features by transposed convolution, and finally obtains the generated sinogram. Specifically, the encoder includes one Padding-7×7 Convolution (Conv)-Instance Norm (IN)-ReLU layer and two 3×3 Conv-IN-ReLU layers. In the first layer, the parameter of padding is 3, and there are 64 filters with a stride of one. The next two layers consist of 128 and 256 filters with a stride of two. In the transformation module, each ResNet block contains two convolution layers with 256 filters on both layers and the parameter of padding is 1. The decoder includes two 3×3 Deconvolution-IN-ReLU layers and one Padding-7×7 Conv-Tanh layer. The two deconvolution layers consist of 128 and 64 filters with a stride of two. In the last layer, the parameter of padding is 3, and it consists of one filter with a stride of one.

2) **Relativistic Discriminator**: The structure of the relativistic discriminator is shown in Figure 4, which is different from the standard discriminator D in Cycle-GAN [28]. The relativistic discriminator predicts the probability that a real sinogram is relatively more realistic than a fake sinogram. The inputs of the relativistic discriminator are pairs of the real sinogram and the generated sinogram. The relativistic discriminator has five layers. The first layer is composed of a 4×4 Conv layer and leaky rectified linear units (Leaky ReLU) and consists of 64 filters with a stride of two. The second and third layers are 4×4 Conv-IN-LeakyReLU layers and consist of 128 and 256 filters, respectively, with a stride of two. The fourth layer is a 4×4 Conv-IN-LeakyReLU layer and consists of 512 filters with a stride of one. The last layer is a 4 × 4 Conv-sigmoid layer with one filter of stride one. The differences of 64×64 patches are input in a sigmoid function to obtain the one-dimensional output.

*2.4 Image Reconstruction from Generated Sinogram*

We obtain the generated HR sinogram by inputting the LR sinogram into the trained SSR-GAN model. Then, we obtain the reconstructed CT image from the generated HR sinogram using the classic reconstruction algorithms, such as the FBP algorithm [30]. By this way, a relatively high-quality reconstructed image is obtained from an LR sinogram.

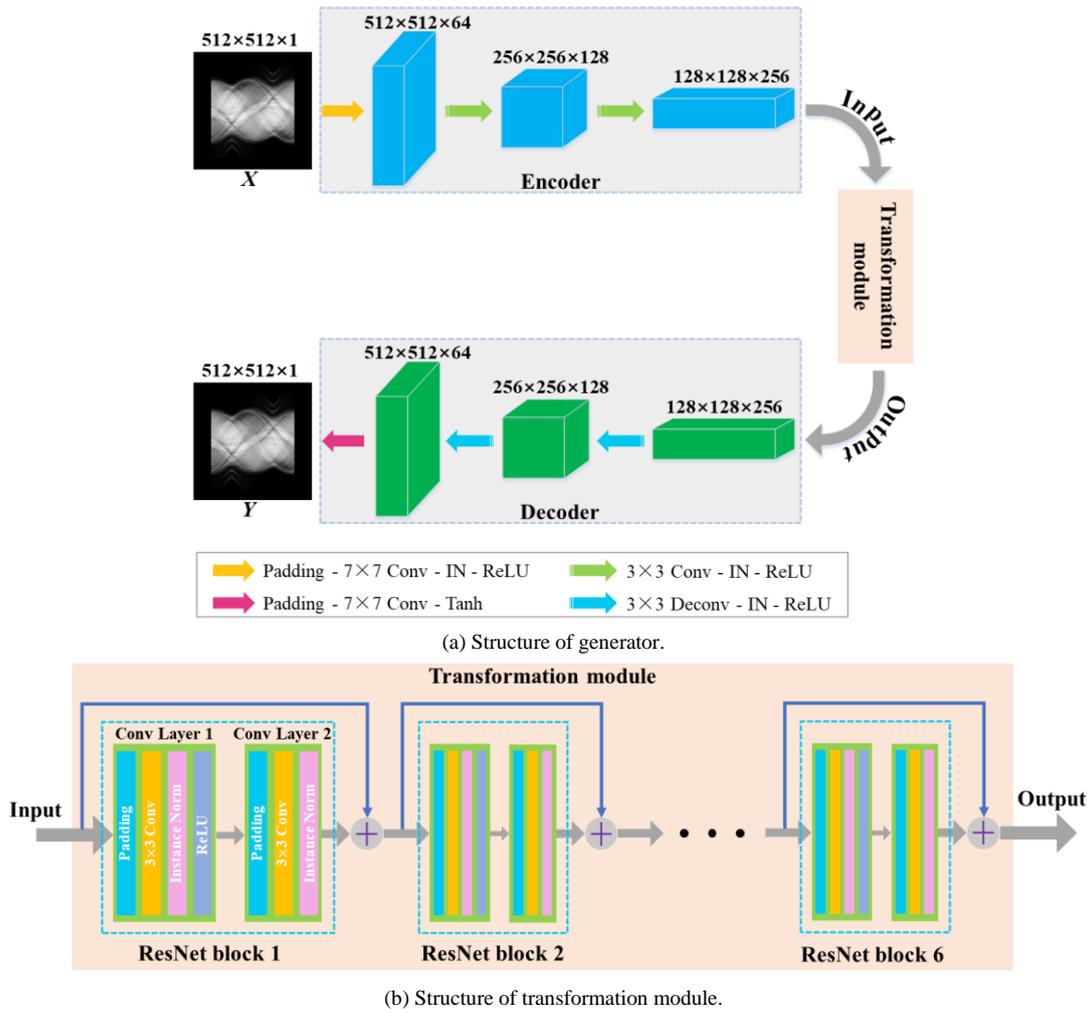

(a) Structure of generator.

(b) Structure of transformation module.

**Figure 3.** Architecture of the sinogram SR generators. In (a), the blue and green boxes indicate the feature vectors obtained by the network layers in the encoder and decoder. $X$ represents the input sinogram, and $Y$ represents the generated sinogram. In (b), there are six ResNet blocks, with the structure of each ResNet block being the same as the others.

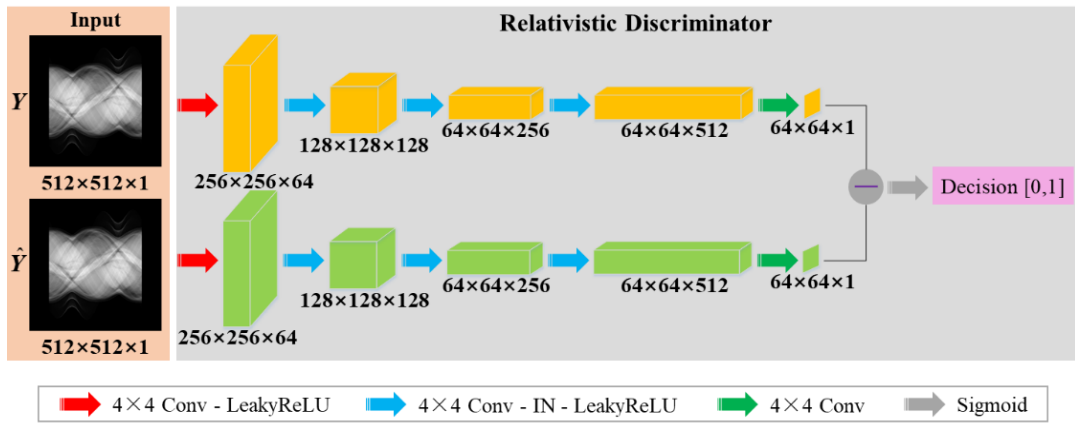

**Figure 4.** Architecture of the discriminators. The yellow and light green boxes indicate the feature vectors obtained by the network layers in the discriminator. $Y$ represents the real sinogram, and $\hat{Y}$ represents the generated sinogram by generator.

## 3. EVALUATION

### 3.1 Experimental Data and Training Details

*3.1.1 Digital CT Image Data.* In our simulated experiment, the experimental datasets were established from a real clinical dataset that contains 2,159 pleural CT images from six patients. The size of image is 512×512. The training data for training the SSR-GAN were prepared with 1,879 CT images from five patients, while the test data were prepared using 280 CT images from the other one patient. The detailed procedure for preparing the training dataset is shown in Table 1.

**Table 1** Establishment of the training dataset

| Procedure for establishing the training dataset |
|---|
| **step 1.** Each CT images was subjected to value normalization. The image value was rescaled to [0, 0.1]. Then, the normalized images were taken as samples for generating sinogram. |
| **step 2.** Siddon's ray-tracing algorithm [39] was applied to simulate the fan-beam geometry. The 1×1 acquisition mode was used to generate the HR sinograms for 360 views in 360° with 512 linear detectors. The size of the generated HR sinograms is 360×512. Then, we added two 76×512 all-zero matrices to the generated sinograms to obtain 512×512 sinograms as the labels of the SSR-GAN. |
| **step 3.** To simulate the 2×2 acquisition mode, we retained the other parameters of the sinograms generated in step 2 and doubled the size of the detector. Then, we generated the LR sinograms for 360 views in 360° with 256 linear detectors. The size of the generated LR sinograms is 360×256. Next, we used the bicubic interpolation algorithm to obtain the 360×512 sinograms from the generated 360×256 sinograms, and we added two 76×512 all-zero matrices to obtain 512×512 sinograms as the inputs of the SSR-GAN. In addition, we added noise to illustrate the practicality of the method. The noise is modeled as Gaussian zero-mean and variance $\sigma^2$ [40]: $$g_i \sim normal(0, \sigma^2), \quad (12)$$ where $i$ denotes the pixels in the sinogram data, $g_i$ denotes the measured sinogram with added Gaussian noise, and the noise variance $\sigma^2$ was set to 5×10$^{-6}$. |
| On the basis of the above procedure, 1,879 pairs of input and label sinograms with a size of 512×512 were prepared. |

For the preparation of the test dataset, the 280 CT images were used to generate LR sinograms by the process explained in step 3 in Table 1.

*3.1.2 Anthropomorphic Head Phantom Data.* To verify the potential SR capability of the proposed model for an actual CT system, we investigated a radiological anthropomorphic head phantom study for clinical applications. The phantom is shown in Figure 5, and its specifications are described in the ICRU Report 48 [41]. In the real data experiment, the CT projection data were acquired by a CT scanner, which is mainly composed of the X-ray source (Hawkeye130, Thales) and a flat-panel detector (Thales 4343F). In the data acquisition, we obtained the projection datasets under different parts of the head phantom for training and testing of the SSR-GAN, respectively. The training and test datasets were completely separated; the two closest slices in the two datasets were also separated by 200 slices. The specific scanning parameters are listed in Table 2. Given the significant difference between the simulated data and the actual collected data, we must retrain the SSR-GAN so that it can be applied in the actual CT system.

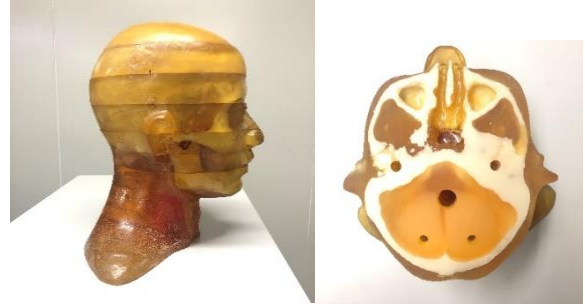

**Figure 5.** Real data experimental phantom: Chengdu Dosimetric Phantom.

**Table 2** Parameters set in the head phantom experiments.

| Parameters | Value of parameter |
|---|---|
| Detector elements | 716 |
| Detector bin size (mm) | 0.592 |
| Distance of source to object (mm) | 529.16 |
| Distance of source to detector (mm) | 998.75 |
| Tube voltage (kVp) | 120 |
| Tube current (μA) | 200 |
| Scanning range (°) | 360 |
| Number of projections | 360 |
| Reconstruction size | 512×512 |

In the experiment, we collected 1002 HR sinograms of phantom slices for training the SSR-GAN. Furthermore, we collected additional 201 HR sinograms of phantom slices for preparing the test dataset. The size of the HR sinograms is 360×512. Limited by experimental conditions, we cannot directly obtain the LR sinograms based on the 2×2 acquisition mode. The 360×256 LR sinograms were obtained by down-sampling the real collected HR sinograms. Note that the value range of all sinograms was normalized to [0, 1] in the data preprocessing stage. For the training datasets, we utilized the bicubic interpolation algorithm to obtain 360×512 sinograms from the LR sinograms, and then added two 76×512 all-zero matrices to get 512×512 sinograms as the inputs of the SSR-GAN. Moreover, we added two 76×512 all-zero matrices to the real collected HR sinograms to obtain 512×512 sinograms as the labels of the SSR-GAN. We obtained 1002 pairs of 512×512 input sinograms and corresponding label sinograms. For testing the SSR-GAN, we obtained the LR sinograms by down-sampling the 201 real collected HR sinograms. Then, we utilized the bicubic interpolation algorithm to obtain 360×512 sinograms from the LR sinograms and added two 76×512 all-zero matrices to generate 512×512 testing sinograms.

*3.1.3 Training Details.* All training and testing procedures were performed using the Pytorch toolbox (ver. 0.4.1) running on an AMAX workstation with two GeForce RTX 2080 Ti GPUs (NVIDIA Corporation). The parameters were updated by the adaptive moment (Adam) optimizer, the learning rate was fixed at 0.0002 in the first 100 epochs and decreased linearly from 0.0002 to 0 over the next 100 epochs, and the exponential decay rates were $\beta_1 = 0.5$ and $\beta_2 = 0.999$. In our experiment, the batch size was 2 in all the training processes. In the training process of the SSR-GAN, approximately 22.205 million parameters were to be learned. The amount of data affects the training time of the SSR-GAN. In the simulated experiment, the time cost was approximately 61 h for the overall training procedure of the SSR-GAN, whereas it was approximately 36 h in the real data experiment. The weight parameter selection of loss function $\mathcal{L}_{SSRGAN}$ is important. A detailed discussion is presented in Section 4.1.

### 3.2 Performance Comparison

In this study, the bicubic interpolation-based [42], U-Net-based [24], and the cycle-GAN-based [29] SR methods were adopted for comparison. We trained the U-Net- and cycle-GAN-based SR methods with different hyperparameter settings to obtain relatively good SR performance. The network structure of the U-Net-based SR method is the original U-Net framework. In addition, the heart of the cycle-GAN-based SR method is the original cycle-GAN framework [28], and we trained the cycle-GAN in the supervised manner. The reconstruction images were obtained from the SR sinograms of different methods by using the FBP algorithm [30].

For the quantitative evaluation of the SR performance of the proposed model, we validated the reconstruction results in terms of three image quality metrics: peak signal-to-noise ratio (PSNR), structural similarity (SSIM) [43], and root mean square error (RMSE). Through extensive experiments, we compared the different SR methods on the simulated and real datasets described in Section 3.1.

## 4. Results

### 4.1 Parameter Selection in Loss Function

In loss function $\mathcal{L}_{SSRGAN}$, the weight parameters $\lambda_1$, $\lambda_2$, and $\lambda_3$ together determine the optimal proportion of different losses in the whole training process of the SSR-GAN. In this study, $\lambda_1$ controls the weight of the cycle-consistency loss. To obtain a relatively good value of parameter $\lambda_1$, we first set $\lambda_2 = 0$ and $\lambda_3 = 0$. It means that the proposed SSR-GAN model was only optimized with the cycle consistency loss and the adversarial loss. In this scheme, we investigated the performance of the SR model on the test dataset. For fairness, each SR model had the same parameter setting, except $\lambda_1$. The 20 LR sinograms from the test dataset were randomly selected for testing the SR performance of the different models. The effect of $\lambda_1$ was quantitatively determined by calculating the average RMSE of the SR sonograms, as shown in Figure 6. From the curves of Figure 6, when $\lambda_1 = 10$, the average RMSE of the SR sinograms reached the minimum. Therefore, we adjusted the other two parameters $\lambda_2$ and $\lambda_3$ based on $\lambda_1 = 10$ to test the effects of the sinogram domain loss and reconstruction image domain loss.

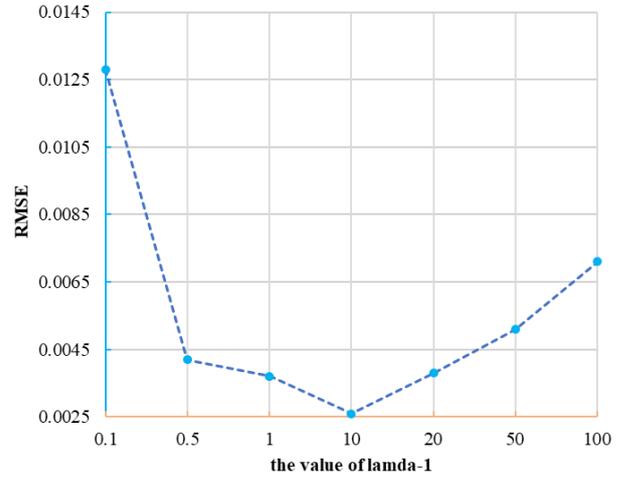

**Figure 6.** Average RMSE of SR sinograms for different values of $\lambda_1$ when $\lambda_2 = 0$ and $\lambda_3 = 0$.

Weight parameters $\lambda_2$ and $\lambda_3$ are important for controlling the balance between the sinogram loss and reconstruction loss. When $\lambda_3$ is extremely small, the proposed model basically cannot benefit from the image domain information. Similarly, when $\lambda_2$ is extremely small, the constraint of the model on the output sinograms is reduced, which is not conducive to the generation of accurate HR sinograms. In our study, we also investigated the performance of the retrained SR model with different values of parameters $\lambda_2$ and $\lambda_3$ on the test dataset. The effect of $\lambda_2$ and $\lambda_3$ were quantitatively determined by calculating the average RMSE of the SR sinograms, as shown in Figure 7. The color of squares indicate that when $\lambda_2 = 1 \times 10^{-4}$ and $\lambda_3 = 500$, the RMSE of the SR sinograms reached the minimum. In summary, we set $\lambda_1 = 10$, $\lambda_2 = 1 \times 10^{-4}$ and $\lambda_3 = 500$ in the experiments.

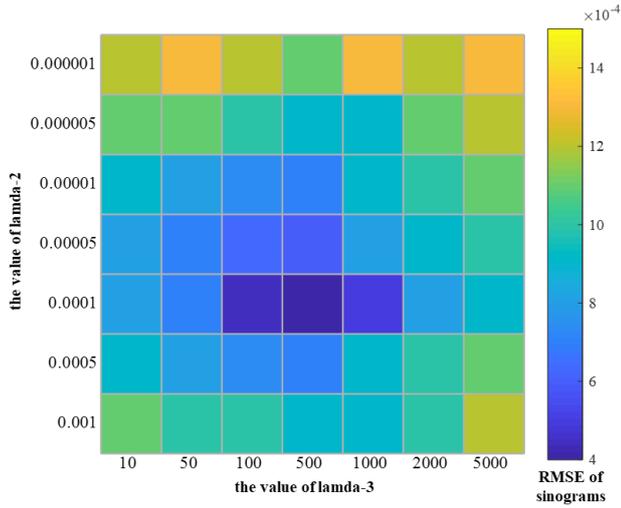

**Figure 7.** Average RMSE of SR sinograms for different values of $\lambda_2$ and $\lambda_3$ when $\lambda_1 = 10$. The darker the color of the square, the smaller the estimated RMSE of the sinograms.

To analyze the sinogram SR capability of the proposed SSR-GAN, four representative sinograms are shown in Figure 8, and the reconstructed CT images of the first and fourth sinograms are shown in Figure 9.

Figure 8 indicates that the proposed SSR-GAN captures more sinogram information. Compared with the HR sinograms obtained under the 1×1 acquisition mode, the sinograms generated by the SSR-GAN only have few errors. Compared with other SR methods, the results obtained by the proposed SSR-GAN had greater accuracy. Figure 9 shows the reconstructed images from the SR sinograms obtained by different SR methods and the error maps between the ground truth and the reconstructed images. Moreover, we selected the two regions of these two slices for further analysis of the reconstructed details, which are shown in the third and sixth rows in Figure 9. We observe that the reconstructed CT images from the LR sinograms were of relatively poor quality, and most of the image details were flooded by artifacts. Compared with other SR methods, the proposed SSR-GAN model showed excellent capabilities in image detail restoration. Based on the proposed model, we can obtain a reconstruction result that is similar to the ground

## 4.2 Simulated Data Experiments

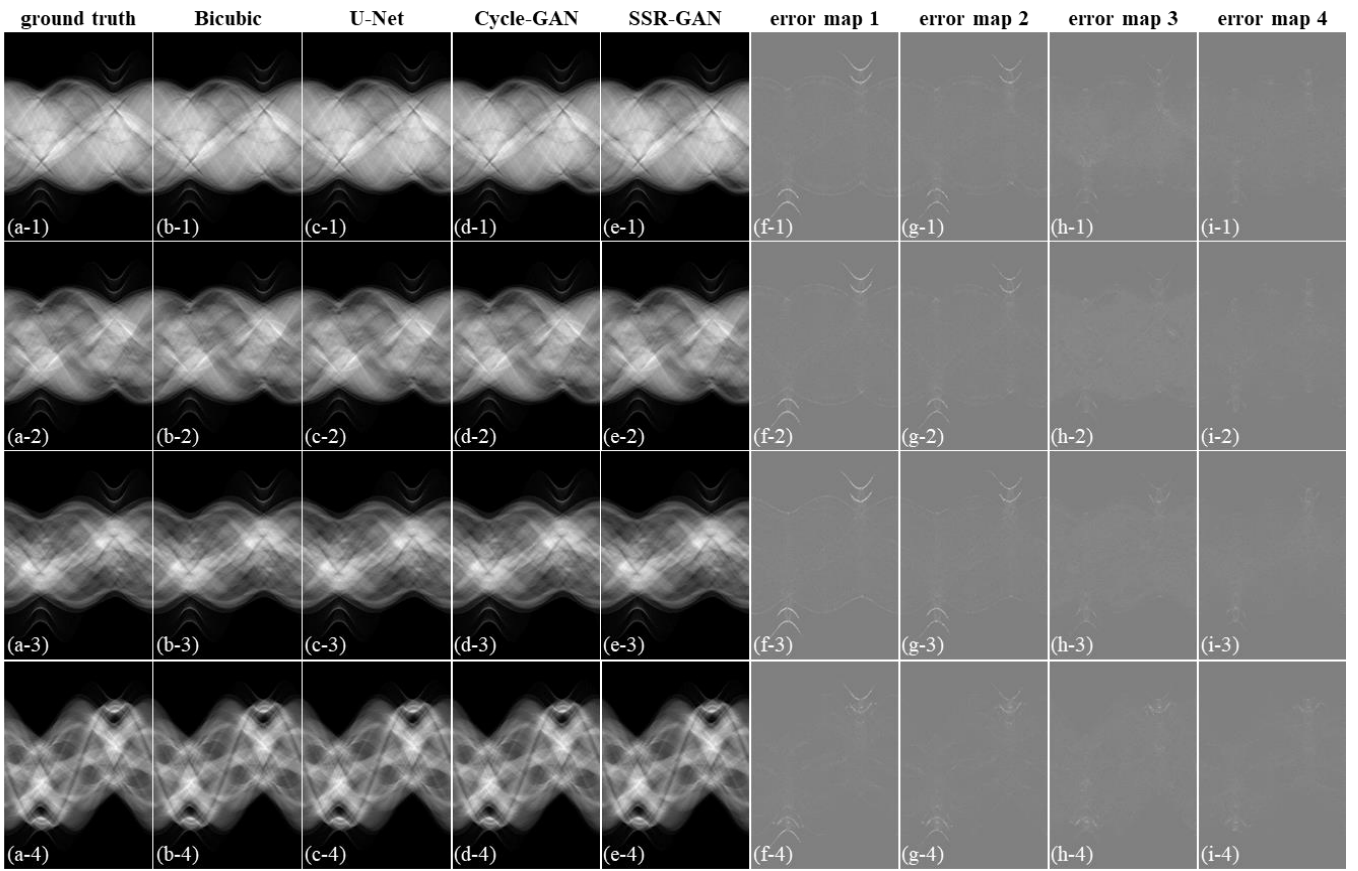

**Figure 8.** SR results for four representative sinograms obtained by different methods. (**a**) generated HR sinogram data under the 1×1 acquisition mode as the ground truth; (**b**) estimated SR sinograms by bicubic interpolation; (**c**) estimated SR sinograms by the U-Net; (**d**) estimated SR sinograms by the cycle-GAN; (**e**) estimated SR sinograms by the proposed SSR-GAN; (**f**) error maps of (**a**,**b**); (**g**) error maps of (**a**,**c**); (**h**) error maps of (**a**,**d**); (**i**) error maps of (**a**,**e**). The display window of (**a**–**e**) is [0, 1]. The display window of (**f**–**i**) is [-0.15, 0.15].

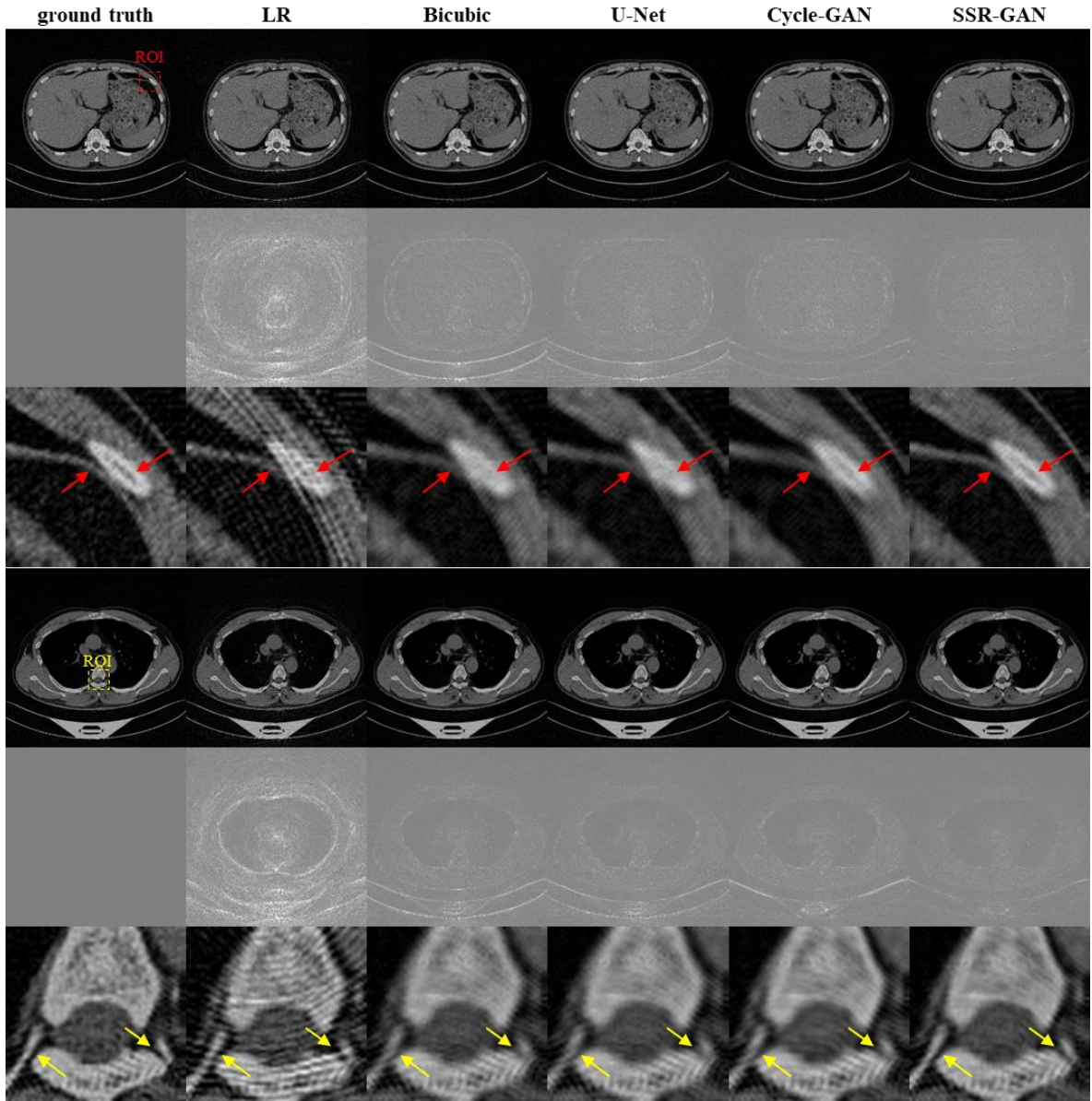

**Figure 9.** Reconstruction results of the first and fourth sinograms. The first row represents the reconstructed images from the HR sinograms, the LR sinograms, and the different SR sinograms. The second row represents the error maps between the ground truth and the resulting images. The third row represents the zoom-in regions of interest of the reconstruction results. The fourth to sixth rows represent the similar results to the first to third rows, which are based on another sinogram. From left to right, the columns represent the results based on the HR sinogram, LR sinogram, SR sinogram by bicubic interpolation, SR sinogram by U-Net, SR sinogram by cycle-GAN, and SR sinogram by the proposed SSR-GAN. The display windows of all figures in the second and fifth rows are [-0.04, 0.04]. The display windows of all figures in the first, third, fourth, and sixth rows are [0, 0.1].

truth. The error maps indicate that the reconstruction results of the proposed model have the smallest difference. Moreover, in the regions of interest (ROIs), we can clearly see that the results based on LR sinograms contained many artifacts. The bicubic interpolation-, U-Net-, and cycle-GAN-based SR methods could depress the artifacts to some extent and recover the images with relatively high quality. However, the image details indicated by arrows are still blurred. Compared with the other methods, the propose method recovered relatively clear image details in the ROIs. This indicates that the proposed method has advantages in detail restoration and blur reduction.

Table 3 shows the quantitative comparison results of the reconstruction images, which can further reflect the performance of the proposed method. The evaluation items indicate the benefit of different SR methods in the image reconstruction of LR sinograms. Among them, the SSR-GAN performed the best in terms of SSIM, which indicates that the results based on SSR-GAN are the most structurally similar to the ground truth. In terms of PSNR and RMSE, the proposed method also exhibited satisfactory performance, indicating that the noise removal is relatively effective. Overall, the quantitative evaluation results of RMSE, PSNR, and SSIM indicate that the proposed method can recover high-quality CT images from LR sinograms.

**Table 3** Quantitative evaluations of reconstruction results based on LR and different SR sinograms (100 testing images).

|  | LR | Bicubic | U-Net | Cycle-GAN | SSR-GAN |
|---|---|---|---|---|---|
| avg. PSNR | 21.3246 | 27.4427 | 28.3096 | 29.5277 | **30.7294** |
| avg. RMSE | $2.4666\times10^{-3}$ | $1.2084\times10^{-3}$ | $1.0286\times10^{-3}$ | $6.7854\times10^{-4}$ | $\mathbf{5.6748\times10^{-4}}$ |
| avg. SSIM | 0.9503 | 0.9830 | 0.9861 | 0.9905 | **0.9925** |

## 4.3 Real Data Experiments

The performance of the proposed method was further evaluated on a radiological anthropomorphic head phantom. We selected two test slices for exhibition and the reconstruction images of these slices are shown at an enlarged field of view with 360×360 square pixels. Figures 10(a) and (b) show the CT images of two representative test slices reconstructed by the FBP algorithm, which were considered as the ground truth of the reconstructed images. In addition, we selected two ROIs to evaluate the ability of different methods to restore image details.

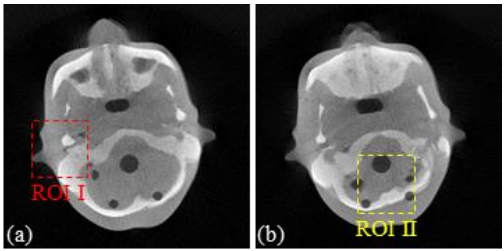

**Figure 10.** Reconstruction results of two representative test slices. (**a**) and (**b**) are slices 1 and 2 reconstructed from the HR sinograms under the 1×1 acquisition mode. The red rectangular box is the ROI of slice 1 and the yellow rectangular box is the ROI of slice 2. The display window is [0, 0.01].

Figure 11 shows the sinograms estimated by the different SR methods and their error maps. Compared with the real HR sinograms, the results of the proposed method only have few errors. The error maps also indicate that the proposed method obtains the most similar results to the collected HR sinograms.

Figure 12 shows the reconstruction results of the head phantom by the FBP algorithm based on the LR sinograms, HR sinograms, and the different SR sinograms. Their error maps are shown in the second and fifth rows of Figure 12. The zoomed-in ROI images are used to reveal texture details, which are shown in the third and sixth rows of Figure 12. Note that the artifacts are obvious in the ROIs of reconstruction images obtained on the basis of LR sinograms. Although the bicubic interpolation-based SR method could depress the artifacts to some extent, the details of reconstruction images were blurred. The U-Net- and cycle-GAN-based SR methods could further improve the image quality; however, the image details indicated by arrows are still not sufficiently clear. Compared with the other SR methods, the proposed method efficiently suppressed the artifacts and recovered the image with high quality. The error maps also indicate that the proposed method had the smallest difference to the ground truth among the four SR methods.

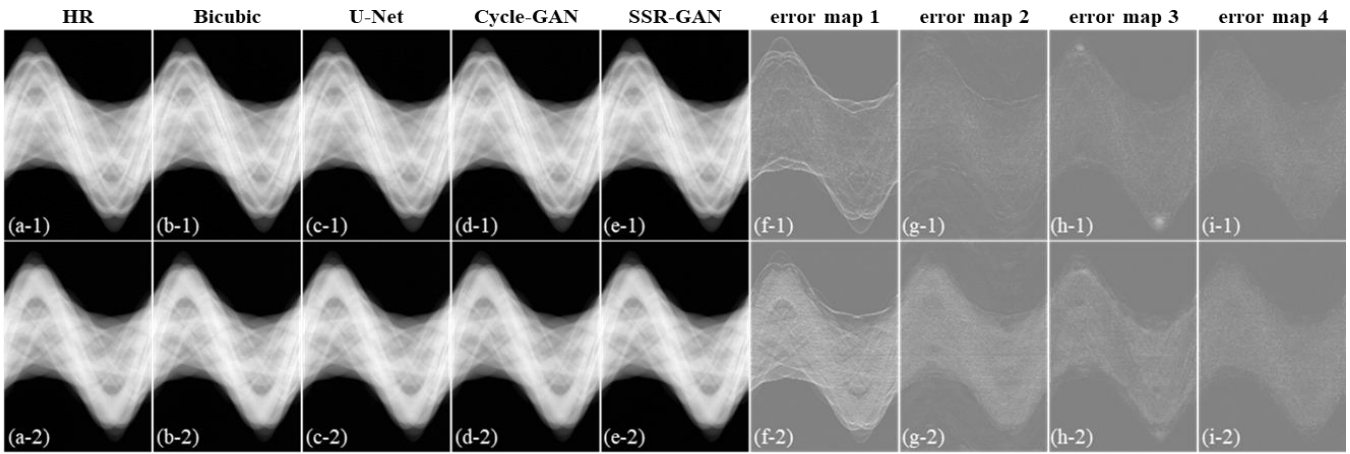

**Figure 11.** SR results for two test sinograms of the head phantom by different methods. (**a**) Collected HR sinogram data under the 1×1 acquisition mode as the ground truth; (**b**) estimated SR sinograms by bicubic interpolation; (**c**) estimated SR sinograms by the classical U-Net; (**d**) estimated SR sinograms by the cycle-GAN; (**e**) estimated SR sinograms by the proposed SSR-GAN; (**f**) error maps of (**a**,**b**); (**g**) error maps of (**a**,**c**); (**h**) error maps of (**a**,**d**); (**i**) error maps of (**a**,**e**). The display window of (**a**–**e**) is [0, 1]. The display window of (**f**–**i**) is [-0.06, 0.06].

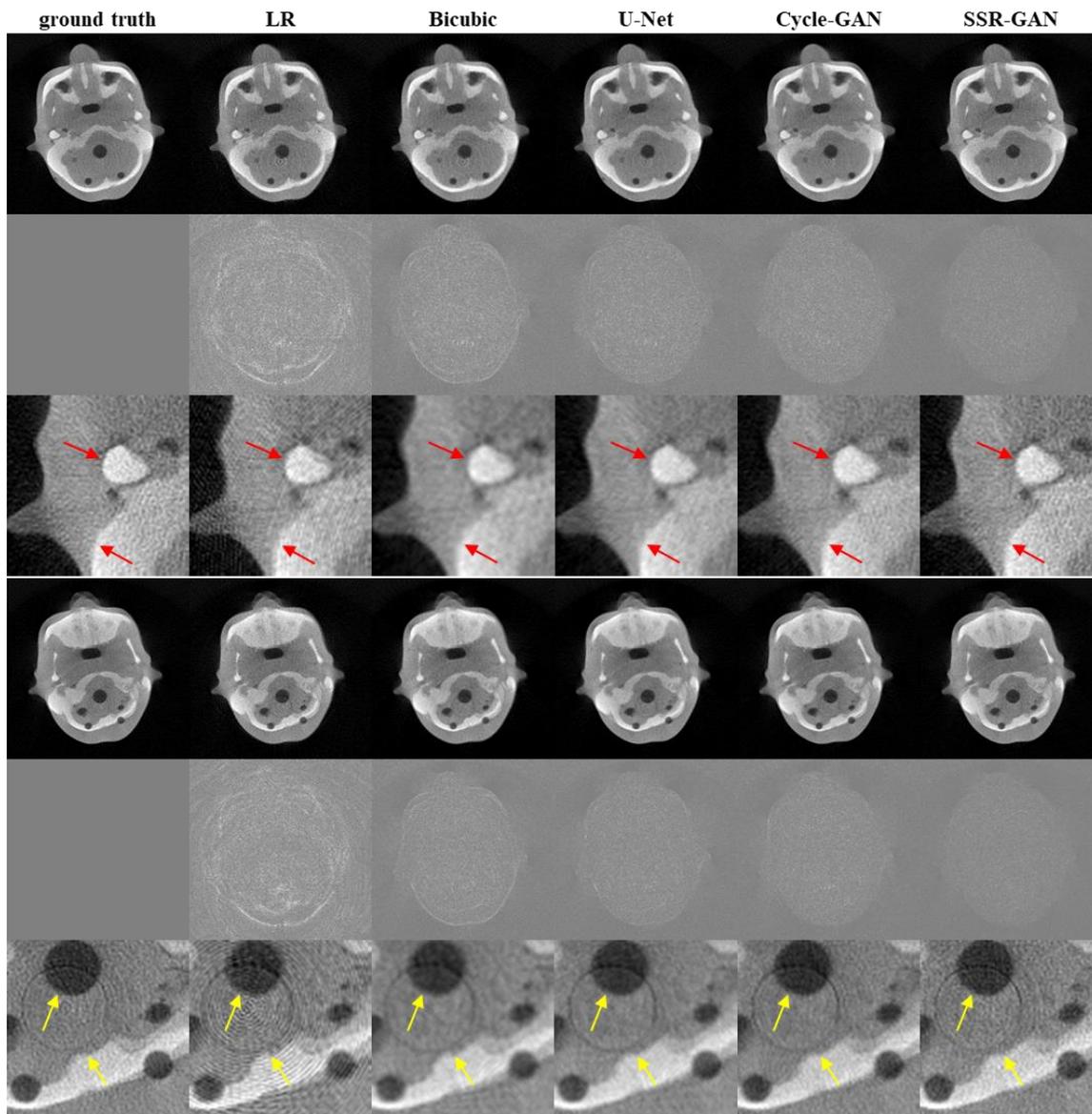

**Figure 12.** Reconstruction results of the anthropomorphic head phantom. The first row represents the reconstructed images from the collected HR sinograms, LR sinograms, and different SR sinograms. The second row represents the error maps between the ground truth and the resulting images. The third row represents the zoom-in ROIs of reconstruction results. The fourth to sixth rows represent the similar results to the first to third rows. The results of the first to third rows are related to slice 1, and those of the fourth to sixth rows are related to slice 2. From left to right, the columns represent the results based on the collected HR sinograms, LR sinograms, SR sinograms by bicubic interpolation, SR sinograms by U-Net, SR sinograms by cycle-GAN, and SR sinograms by the proposed SSR-GAN. The display windows of all figures in the second and fifth rows are [-0.003, 0.003]. The display windows of all figures in the first, third, fourth, and sixth rows are [0, 0.01].

**Table 4** Quantitative evaluations of reconstruction results based on LR and different SR sinograms in the anthropomorphic head study (20 testing images).

|           | LR | Bicubic | U-Net | Cycle-GAN | SSR-GAN |
|---|---|---|---|---|---|
| avg. PSNR | 25.2521 | 29.9985 | 31.0190 | 31.4205 | **31.7877** |
| avg. RMSE | $1.1749 \times 10^{-4}$ | $5.7586 \times 10^{-5}$ | $5.3408 \times 10^{-5}$ | $3.8338 \times 10^{-5}$ | $\mathbf{3.6789 \times 10^{-5}}$ |
| avg. SSIM | 0.9965 | 0.9989 | 0.9992 | 0.9994 | **0.9995** |

The quantitative evaluation results of reconstructed images are shown in Table 4. Results show that the proposed method exhibits good performance in terms of noise suppression and structure preservation; these results are consistent with the findings given in Table 3. Thus, the evaluation of the head phantom also validates the efficiency of the proposed method in CT imaging under the 2×2 acquisition mode.

## 5. Discussion and conclusion

In CT imaging, the 2×2 acquisition mode can improve the collection efficiency of the projection and significantly reduce the X-ray exposure time. However, this acquisition mode is rarely adopted because of the low-quality reconstruction results. The study of sinogram SR problem is beneficial to improving the reconstruction quality of CT images under the 2×2 acquisition mode. Currently, most of the existing SR methods only focus on the image domain, and the information of the projection domain has not received great attention, thus limiting the application of this acquisition mode in practice. A feasible approach to increase the effective information of LR sinograms is to use sinogram SR methods for image reconstruction. Inspired by the recent successful image-to-image translation applications of the cycle-GAN, we proposed the SSR-GAN for supervised generation of SR sinograms. To ensure that the generated SR sinograms have high accuracy, we designed the relativistic discriminator to predict the probability that a real sinogram is relatively more realistic than a fake sinogram. Moreover, we added the weighted sinogram domain loss and reconstruction image domain loss for the proposed SSR-GAN, which can increase the fidelity of the reconstructed image. The experiment results indicate that the proposed method outperforms other SR methods.

For the proposed SSR-GAN, appropriate hyperparameters are the guarantee of network performance. In particular, the different weights in the total loss function significantly affect network training. By adjusting the weights, the different losses are relatively balanced, and therefore, the network can realize the continuous interaction of the projection domain and image domain information during training. Thus, the generated HR projection has high accuracy. The selection of different weights requires considerable experimental arguments, which can be seen in Section 4.1. The experiment results indicated that additional training was needed for the sinogram data collected by an actual CT system, probably due to the significant difference in the distribution of simulated data and real data. Obviously, the difference in data distribution also exists in different CT imaging systems. However, we can collect the training data from the different CT imaging systems, and then train multiple models for practical application. In addition, for a specific CT system, we can collect the projection data of different scanned objects to expand the training dataset. The robustness of the trained network model is enhanced with the increase in the collected projection data.

Although the proposed model has achieved compelling SR results, there still exist some limitations. First, compared with the general DL-based methods, a considerably longer training time was required by the proposed SSR-GAN. In this aspect, we should consider designing more efficient architecture that would render the network capable of learning more complex features of sinograms with less computational cost and lower model complexity in the future. Second, although the proposed method can recover image details with high quality, some subtle features may not be always faithfully recovered. To address this issue, future work will explore an improved reconstruction algorithm that uses the SR sinogram as the initial iteration value and utilize the scanned LR sinogram to construct the fidelity term. The improved reconstruction algorithm may will further enhance the image details.

In conclusion, this study proposed an efficient method to obtain HR sinograms from LR sinograms under the 2×2 acquisition mode by taking advantage of the neural network in nonlinear mapping and utilizing the information interaction between the projection domain and the image domain. Based on the HR sinograms obtained by the proposed SSR-GAN, relatively high-quality reconstruction images can be obtained by the FBP algorithm. The proposed method effectively promotes the practical application of the 2×2 acquisition mode and has considerable potential for reducing the X-ray radiation dose of CT imaging, increasing the CT scanning speed and reducing the CT production cost.

## Acknowledgements

This work was supported by the National Natural Science Foundation of China (Grant No. 61601518) and the National Science Foundation for Post-doctoral Scientists of China (Grant No. 2019M663996).